\documentstyle[aasms4]{article}
\begin{document}
\title{A Survey for Low Surface Brightness Galaxies Around
M31.\ II.\ The Newly Discovered Dwarf Andromeda VI}
\author{Taft E.\ Armandroff}
\affil{Kitt Peak National Observatory, National Optical Astronomy
Observatories\altaffilmark{1},\\ P.O.\ Box 26732, Tucson, AZ 85726}
\authoremail{armand@noao.edu}
\author{George H.\ Jacoby\altaffilmark{2}}
\affil{Kitt Peak National Observatory, National Optical Astronomy
Observatories\altaffilmark{1},\\ P.O.\ Box 26732, Tucson, AZ 85726}
\authoremail{jacoby@noao.edu}
\and
\author{James E.\ Davies\altaffilmark{3}}
\affil{Kitt Peak National Observatory, National Optical Astronomy
Observatories\altaffilmark{1},\\ P.O.\ Box 26732, Tucson, AZ 85726;\\
and\\ Department of Astronomy, University of Wisconsin, Madison,
WI 53706}
\authoremail{jdavies@eta.pha.jhu.edu}
\altaffiltext{1}{The National Optical Astronomy Observatories are
operated by the Association of Universities for Research in
Astronomy, Inc., under cooperative agreement with the National
Science Foundation.}
\altaffiltext{2}{Visiting Astronomer, Steward Observatory, University
of Arizona.}
\altaffiltext{3}{Current address: Department of Physics and Astronomy,
Johns Hopkins University, Baltimore, MD 21218.}
\slugcomment{Accepted for publication in The Astronomical Journal
(September 1999 issue)}
\lefthead{Armandroff, Jacoby \& Davies}
\righthead{Low Surface Brightness Galaxies Around M31}

\begin{abstract}
We present $B$-, $V$-, and $I$-band images, as well as an H$\alpha$
image, of And VI. This is the second newly identified dwarf spheroidal
(dSph) companion to M31 found using a digital filtering technique
applied to the second Palomar Sky Survey for which $\sim$1550 deg$^2$
now have been surveyed. And VI was confirmed to be a nearby dSph galaxy
when it resolved into stars easily with a short 4-m $V$-band exposure.
Sub-arcsec images taken at the Kitt Peak WIYN 3.5-m telescope provided
($I$,$V$--$I$) and ($V$,$B$--$V$) color--magnitude diagrams that yield
a distance of 775 $\pm$ 35 kpc using the tip of the red giant branch
method, and a mean metallicity of [Fe/H] $= -1.58\pm0.20$ with a
dispersion of $\sim0.3$ dex.  And VI has a galactocentric distance of
$\sim271$ kpc and $M_V = -11.3$. All observed properties of And VI are
consistent with its classification as a dSph companion to M31. Despite
the recent identification of the And V, And VI, and Cas dwarfs, the
Local Group luminosity function remains highly deficient in faint
galaxies relative both to CDM simulations of its formation, and to the
luminosity functions for richer clusters of galaxies.
\end{abstract}

\section{Introduction}

\begin{figure}[p]
\epsscale{0.85}
\plotone{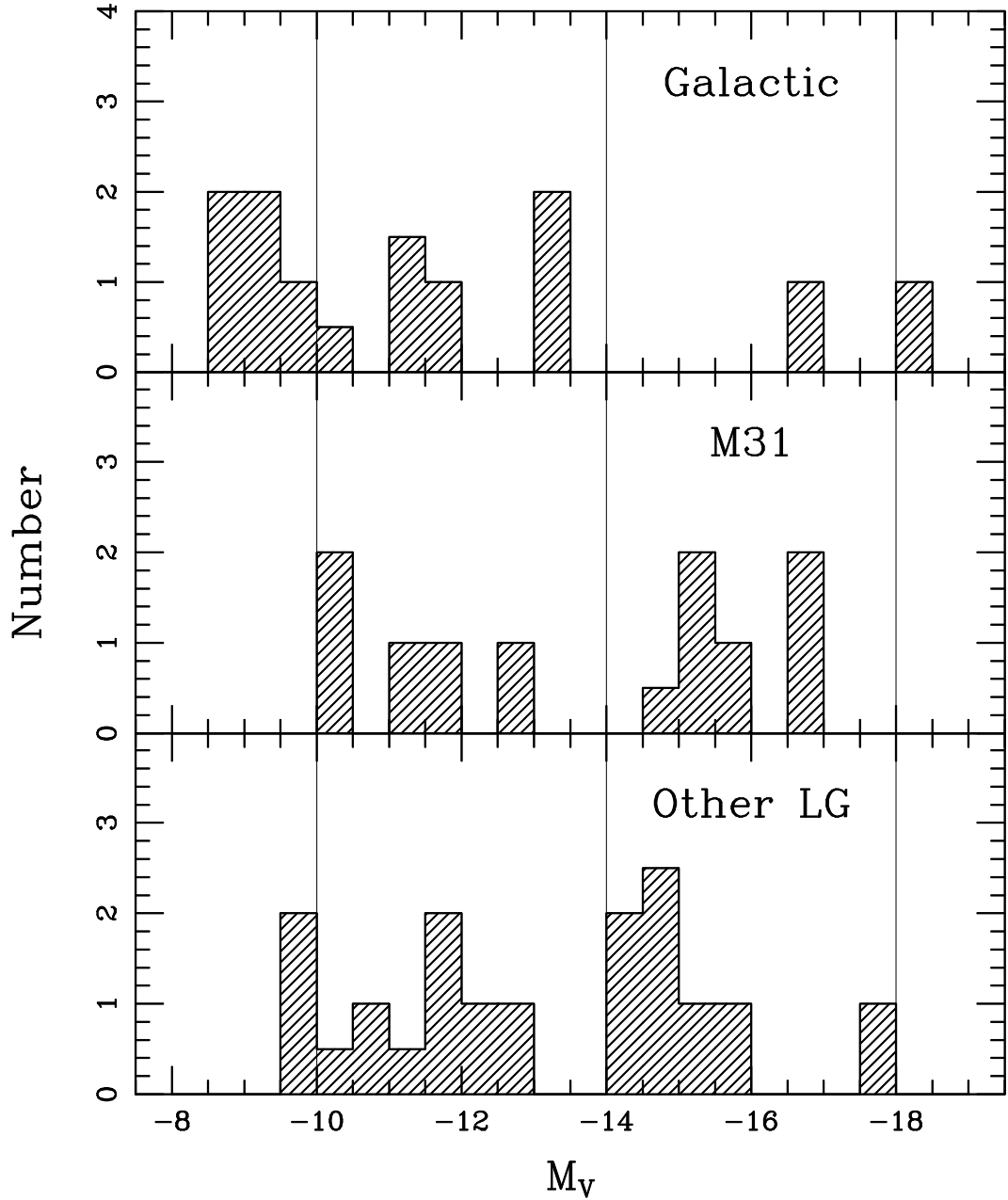}
\caption{Luminosity functions for Local Group dwarf galaxies ($M_V$
values from Table 4 of Mateo 1998).  Those dwarfs associated with the
Galaxy, M31 and the general Local Group are shown separately.  Note
that the faintest dwarf galaxies appear to be more common around the
Galaxy than in other parts of the Local Group, although this could be
caused by increasing incompleteness at faint $M_V$ with increasing
distance.  The Kolmogorov-Smirnov test finds the Galactic $M_V$
distribution to be different from that of M31 at the 99.0\% confidence
level and from that of the general Local Group at the 98.8\% confidence
level.}
\label{gal_lum_fun}
\end{figure}

The properties, frequency, and origin of dwarf galaxies have important
implications for galaxy formation and cosmology.  With an eye toward
these implications, Gallagher \& Wyse (1994), Grebel (1997), Da Costa
(1998), and Mateo (1998) recently have reviewed the dwarf galaxies of
the Local Group.  As an example of a property of this class of galaxy
where our knowledge is far from complete, consider the faint end of
the galaxy luminosity function.  Figure \ref{gal_lum_fun} shows a
luminosity function, based on Table 4 of Mateo (1998), for all dwarf
galaxies in the Local Group, with those dwarfs associated with the
Milky Way, M31, and the general Local Group indicated.\footnote{See
Table 1 of Mateo (1998) for the association between individual galaxies
and subgroups within the Local Group.  For the few galaxies with two
possible associations, we count them as ``half galaxies'' in each
subgroup.  The absolute magnitudes of the LMC and SMC are from
Westerlund (1997).}  Note that the faintest dwarf galaxies appear to be
more common around the Milky Way than in other parts of the Local
Group.  Is this a real effect or, instead, due to increasing
incompleteness for low-luminosity dwarf galaxies as one looks beyond
the outer halo of the Galaxy?  The only way to resolve this and related
questions is through more sensitive searches for low-luminosity
galaxies in the Local Group.
\placefigure{gal_lum_fun}

We have undertaken a survey for low surface brightness dwarf galaxies
around M31.  Our methodology is the application of digital filtering
techniques to the Second Palomar Sky Survey (POSS-II; Reid et
al.\ 1991, Reid \& Djorgovski 1993).  Our survey seeks to cover a
larger area around M31 and to reach a deeper limiting central surface
brightness than the pioneering survey of van den Bergh (1972, 1974),
which revealed three M31 dwarf companions (And I, II \& III).  In the
first paper of this series (Armandroff, Davies, \& Jacoby 1998b;
hereafter referred to as Paper I), we described our methodology,
discussed its application to 1550 deg$^2$ of the sky survey, and
announced the first discovery of this survey: the new dwarf galaxy And
V.  We also presented a color--magnitude diagram for And V which yields
a distance to this galaxy that is equivalent, within the uncertainties,
to that of M31.  The line-of-sight distance to And V and its projected
distance from M31 suggest association with M31.  The color--magnitude
diagram also implies a mean metal abundance for And V of [Fe/H]
$\approx$ --1.5 and lacks bright blue stars.  This absence of a young
population in the color--magnitude diagram, the lack of H$\alpha$
emission in the differencing of our deep H$\alpha$ and $R$ images, and
the non-detection of far-infrared emission all imply that And V is a
dwarf spheroidal (dSph) galaxy, and not a dwarf irregular galaxy.  And
V thus joined M31's three other dwarf spheroidal companions (And I, II
\& III) and three more luminous dwarf elliptical companions (NGC 147,
185 and 205).  IC 10, a dwarf irregular, LGS 3, a transition object
between dwarf irregular and dwarf spheroidal, and M32, a low-luminosity
elliptical, complete the inventory of dwarf galaxies commonly accepted
as associated with M31.

In this paper, we discuss the next product of our survey: And VI.  The
newly discovered galaxy And VI is found to be another dwarf spheroidal
companion to M31 (see also Karachentsev \& Karachentseva 1999; Grebel
\& Guhathakurta 1999; Hopp et al.\ 1999).

\section{Finding And VI}

Using the methodology of Paper I, we found an excellent low surface
brightness dwarf candidate at the following celestial coordinates:
$\alpha$ = 23$^{\rm h}$51$^{\rm m}$46.3$^{\rm s}$, $\delta$ =
+24$^\circ$34$^\prime$57$^{\prime\prime}$ (J2000.0).  These correspond
to Galactic coordinates of: $l$ = 106.0$^\circ$, $b$ = --36.3$^\circ$.
Anticipating the evidence presented below and following van den Bergh
(1972), we will call this new galaxy Andromeda VI.  Figure
\ref{survey_map} shows And VI's location on the sky relative to M31,
M31's known companions, and our search area.  Figure \ref{and6dss}
displays And VI on the digitized POSS-II, raw and after our enhancement
procedure.  One sees a strong resemblance between And VI and the known
M31 dwarf spheroidals And II \& III on the raw and processed POSS-II
images (compare Figure \ref{and6dss} with Figure 1 of Armandroff et
al.\ 1998b).  And VI is visible on the POSS-II $B$ and $R$
transparencies. And I, II \& III are also visible on the POSS-II
transparencies.
\placefigure{survey_map}
\placefigure{and6dss}

There are three possible reasons why And VI was not recognized until
recently.  First, it is just outside the van den Bergh (1972, 1974)
search area.  Second, And VI is located right next to a bright star, so
it might have been thought to be a ghost reflection caused by that
star.  Finally, examination of the POSS-II transparencies reveals
diffuse Galactic emission to be common in this region, so And VI could
have been interpreted as part of this extended emission.

\section{Follow-Up Observations}

As a first step toward clarifying its nature, And VI was imaged at the
Kitt Peak National Observatory 4-m telescope prime focus with the T2KB
CCD on 1998 January 23 (U.T.).  Only one $V$ exposure of 300 sec was
obtained due to And VI's limited visibility in January.  And VI
resolved nicely into stars in this short $V$ image, suggesting that it
is indeed a nearby dwarf galaxy.\footnote{Several months after our
confirmation that And VI is a bona fide nearby dwarf galaxy, we
reported our discovery in the ``Dwarf Tales'' Newsletter (Armandroff et
al.\ 1998c).  Karachentsev \& Karachentseva (1998) also reported their
independent finding of And VI, calling it the Pegasus Dwarf, in ``Dwarf
Tales'' and in a subsequent journal submission (Karachentsev \&
Karachentseva 1999).}

And VI was imaged more deeply using the WIYN\footnote{The WIYN
Observatory is a joint facility of the University of Wisconsin-Madison,
Indiana University, Yale University, and the National Optical Astronomy
Observatories.} 3.5-m telescope, the S2KB CCD (a 2048 $\times$ 2048
STIS device) mounted at the Nasmyth focus, and the $B$, $V$ and $I$
filters on several occasions during 1998 September and October.  The
scale was 0.197 arcsec/pixel, and each image covered 6.7$\times$6.7
arcmin.  The exposure times, dates of observation, and the resulting
image quality are listed in Table \ref{obs_table}.  The deep images
were taken during three nights when the seeing conditions were
excellent (0.6 -- 0.8 arcsec FWHM).  These deep exposures were
calibrated via images taken on a photometric night during which 13
Landolt (1992) standards were observed.  All images were overscan
corrected, bias subtracted, and flat fielded in the standard manner.
The $I$ images were defringed using the scaled combination of several
dark-sky fringe frames.

Figure \ref{and6_ccd_image} displays a color image of And VI made from
the $B$, $V$, and $I$ CCD images from WIYN.  And VI resolves nicely
into stars on the $B$, $V$ and $I$ images.  The And VI images exhibit a
smooth stellar distribution and a resemblance with the other M31 dwarf
spheroidals (compare Figure \ref{and6_ccd_image} with Figure 1 of
Armandroff et al.\ 1993 and/or Figure 4 of Paper I).  And VI does not
look lumpy or show obvious regions of star formation that would suggest
a dwarf irregular, as opposed to dwarf spheroidal, classification.  No
obvious globular clusters are seen in And VI, though none are expected
for a low-luminosity dwarf spheroidal (e.g., Sec.\ 4.3 of Da Costa \&
Armandroff 1995).
\placefigure{and6_ccd_image}

In order to look for possible ionized gas, And VI was observed with the
KPNO 4-m telescope in H$\alpha$ narrowband and $R$ on 1998 July 15
(U.T.) with the CCD Mosaic Imager (Armandroff et al.\ 1998a).  The
H$\alpha$ filter used has a central wavelength of 6569 \AA\ and a width
of 80 \AA\ FWHM in the f/3.2 beam of the telescope.  Three narrowband
exposures of 700 sec each and three $R$ exposures of 300 sec each were
obtained.  The average seeing was 1.2 arcsec FWHM, sampled at 0.26
arcsec/pixel.  We computed scaling ratios between the combined
H$\alpha$ and $R$ images and performed the subtraction. Figure
\ref{and6halpha} shows both the H$\alpha$ image and the
continuum-subtracted H$\alpha$ image.  No diffuse H$\alpha$ emission or
H {\sc ii} regions are detected in the And VI continuum-subtracted
image.  The only features seen in this image are in the vicinity of the
brighter stars, where residuals are caused by the slightly different
behavior of the point spread function (PSF) between $R$ and H$\alpha$.
No diffuse H$\alpha$ emission or H {\sc ii} regions were found in And V
either (Paper I).  The lack of H$\alpha$ emission in And VI rules out
current high-mass star formation.  Almost all Local Group dwarf
irregular galaxies are strongly detected in H$\alpha$ (e.g., Kennicutt
1994).  And VI's lack of H$\alpha$ emission, coupled with its smooth,
regular appearance on broadband images, suggests that it is a dwarf
spheroidal galaxy (rather than a dwarf irregular).
\placefigure{and6halpha}

We also examined the IRAS 12-, 25-, 60-, and 100-micron maps of the And
VI region, using the FRESCO data product from IPAC\footnote{IPAC is
funded by NASA as part of the IRAS extended mission under contract to
JPL.}.  And VI is not detected in any of the IRAS far-infrared bands.
And I, II, III  \& V are not detected by IRAS either (see Paper I).
Far-infrared emission, as seen by IRAS, is the signature of warm dust.
As discussed in Paper I, some Local Group dwarf irregular galaxies are
detected by IRAS.  And VI's non-detection in the far infrared is
consistent with And VI being a dSph galaxy.  However, the non-detection
by IRAS does not eliminate the possibility that And VI is a
low-luminosity relatively quiescent dwarf irregular galaxy.  The
H$\alpha$ data and the color--magnitude diagram provide stronger
evidence on And VI's classification.

\section{Color--Magnitude Diagram}

The next step in our study of And VI is the construction of a
color--magnitude diagram for its resolved stars  in order to derive its
distance and explore its stellar populations characteristics.
Instrumental magnitudes were measured on the deep And VI WIYN images
using the IRAF implementation of the {\sc daophot} crowded-field
photometry program (Stetson 1987, Stetson et al.\ 1990).  The standard
{\sc daophot} procedure was used, culminating in multiple iterations of
{\sc allstar}.  Stars with anomalously large values of the {\sc
daophot} quality-of-fit indicator CHI were deleted from the photometry
lists.

Large-aperture photometry was performed for the 13 well-exposed Landolt
(1992) photometric standard stars.  Adopting mean atmospheric
extinction coefficients for Kitt Peak, photometric transformation
equations were derived.  Only a zeropoint and a linear color term of
small size were needed to represent the transformation for each filter.
For each of $B$, $V$, and $I$, several stars that are well exposed on
both the deep And VI images and the And VI images from the photometric
calibration night were used to derive the magnitude offsets between the
deep images and the standard system.

The ($I$,$V$--$I$) color--magnitude diagram of And VI is shown in
Figure \ref{cmd_fiducial}.  We used the spatial distribution of the
stars in the red-giant region of the color--magnitude diagram to
determine a reasonably precise center for And VI.\footnote{The (x,y)
coordinates of this center were transformed to celestial coordinates
using the Hubble Space Telescope Guide Star Catalog.  This is the
origin of the celestial coordinates for And VI given in the first
paragraph of Sec.\ 2.}  Only stars within 370 pixels (73 arcsec) of
this center are plotted in Figure \ref{cmd_fiducial}.  This radius was
chosen in order to provide a reasonably large sample of And VI members
while minimizing contamination by field stars and is similar to the
core radius of 76 arcsec (Caldwell 1999).  Our And VI color--magnitude
diagram clearly shows a red giant branch.  The contamination of the
color--magnitude diagram by foreground stars appears rather minor (see
also below).
\placefigure{cmd_fiducial}

In order to interpret the color--magnitude diagram, an estimate of the
foreground reddening is necessary.  Note that different authors have
adopted slightly different reddenings to the M31 companions.  We adopt
E($B$--$V$) = 0.06 $\pm$ 0.01 for And VI based on the IRAS/DIRBE
results of Schlegel, Finkbeiner, \& Davis (1998).  This value is 0.02
mag larger than that adopted by Grebel \& Guhathakurta (1999) because
they applied a correction for the mean difference between the Schlegel
et al.\ (1998) E($B$--$V$) values and those of Burstein \& Heiles
(1982).  Assuming $A_V$ = 3.2$\times$E($B$--$V$) and the E($B$--$V$) to
E($V$--$I$) conversion of Dean, Warren, \& Cousins (1978), E($V$--$I$)
= 0.08 $\pm$ 0.01 and $A_I$ = 0.11 $\pm$ 0.02.

A distance can be derived for And VI using the $I$ magnitude of the tip
of the red giant branch (Da Costa \& Armandroff 1990; Lee et
al.\ 1993a).  Figure \ref{and6_lf} shows a cumulative luminosity
function in $I$ for And VI (again, for stars within 370 pixels [73
arcsec] of the And VI center).  Color limits of 0.9 $<$ $V$--$I$ $<$
2.2 have been used in constructing the luminosity function in order to
maximize its sensitivity to And VI red giants and to minimize its
dependence on foreground contamination.  No correction for foreground
contamination has been applied to the luminosity function for the
following reasons.  Even in the corners of our And VI images, the And
VI giant branch is seen.  Thus we have no true foreground/background
measurement.  However, a color--magnitude diagram for the corners of
our images (distance from And VI's center $>$ 191 arcsec) that covers
2.8 times more area than represented in Figures \ref{cmd_fiducial} and
\ref{and6_lf} shows no stars with $19 < I < 21$ within the color limits
used to calculate the luminosity function.  Therefore, over the
magnitude and color intervals that we are identifying the red giant
branch tip, Figures \ref{cmd_fiducial} and \ref{and6_lf} are quite free
of field contamination.  Based on the magnitude at which the luminosity
function begins to rise strongly (see Figure \ref{and6_lf}, especially
the inset which shows the derivative of the cumulative luminosity
function), the $I$ magnitude of the red giant branch tip in And VI is
20.56 $\pm$ 0.10.  This value is supported by the apparent location of
the red giant branch tip in the color--magnitude diagram (see Figure
\ref{cmd_fiducial}).  For metal-poor systems such as And VI (as
demonstrated below), the red giant branch tip occurs at $M_I$ = --4.0
(Da Costa \& Armandroff 1990; Lee et al.\ 1993a).  This results in
distance moduli of ($m$--$M$)$_I$ = 24.56 and ($m$--$M$)$_0$ = 24.45
$\pm$ 0.10, corresponding to a distance of 775 $\pm$ 35 kpc.
\placefigure{and6_lf}

As discussed in Da Costa et al.\ (1996) and Paper I, the best available
Population II distances for M31 (on the same distance scale as the red
giant branch tip calibration used above) are: 760 $\pm$ 45 kpc based on
the red giant branch tip (Mould \& Kristian 1986) or RR Lyrae stars
(Pritchet \& van den Bergh 1988) in the M31 halo; or 850 $\pm$ 20 kpc
from the horizontal branch magnitudes of eight M31 globular clusters
(Fusi Pecci et al.\ 1996).  Our distance for And VI is consistent,
within the combined uncertainties, with both of these M31 distances.
Thus And VI is at approximately the same distance along the line of
sight as M31, suggesting an association between And VI and M31.  Note
also that our line-of-sight distance of 775 $\pm$ 35 kpc for And VI is
very similar to those of And I (810 $\pm$ 30 kpc; Da Costa et
al.\ 1996), And III (760 $\pm$ 70 kpc; Armandroff et al.\ 1993), and
And V (810 $\pm$ 45 kpc; Paper I), all on the same distance scale.  And
VI's projected distance from the center of M31 is 271 kpc.  The
Galactic dwarf spheroidal Leo I has a Galactocentric distance of
$\sim$250 kpc, comparable to And VI's projected distance from M31.  The
above line-of-sight and projected distances strongly suggest that And
VI is a member of M31's satellite system.

Two other groups have recently published color--magnitude diagrams for
And VI (Grebel \& Guhathakurta 1999; Hopp et al.\ 1999).  Their
distances for And VI based on the tip of the red giant branch, 830
$\pm$ 80 kpc (Grebel \& Guhathakurta 1999) and 795 $\pm$ 75 kpc (Hopp
et al.\ 1999), are consistent with our value of 775 $\pm$ 35 kpc.  All
three determinations adopt the same zeropoint for the tip of the red
giant branch method and thus are on the same distance scale.

Based on the ($V$,$B$--$V$) and ($I$,$V$--$I$) color--magnitude
diagrams, we identify the two obvious bright blue stars in Figures
\ref{cmd_fiducial} and \ref{cmd_blue} as possible post-asymptotic giant
branch (PAGB) stars. These two stars have apparent $B$, $V$, and $I$
magnitudes of 21.11, 20.95, 20.13, and 21.44, 21.43, 21.30,
respectively. Bond (1996) suggested that PAGB stars may be useful
Population II distance indicators where $<$$M_V$$>$ = -3.4 $\pm$ 0.2.
The average observed, extinction-corrected, $V$ magnitude for these two
stars is $21.00 \pm 0.24$.  Thus, the PAGB distance modulus to And VI
is ($m$--$M$)$_0$ = 24.40 $\pm$ 0.31, or 760 $\pm$ 110 kpc, in
excellent agreement with distances based on the red giant branch tip.

Using our distance from the red giant branch tip and adopted reddening
for And VI, we have overplotted its color--magnitude diagram with
fiducials representing the red giant branches of Galactic globular
clusters that span a range of metal abundance (Da Costa \& Armandroff
1990).  Figure \ref{cmd_fiducial} displays the And VI ($I$,$V$--$I$)
color--magnitude diagram with fiducials for M15 ([Fe/H] = --2.17), M2
(--1.58), NGC 1851 (--1.16), and 47 Tuc (--0.71, dashed line).  In the
color--magnitude diagram, the bulk of the And VI red giants are bounded
by the M15 and NGC 1851 loci.  The mean locus of the And VI red giant
branch is similar to the M2 fiducial.

For an old stellar system (as demonstrated below for And VI), the
($V$--$I$)$_0$ color of the red giant branch is primarily sensitive to
metal abundance.  Armandroff et al.\ (1993) presented a linear relation
between ($V$--$I$)$_{0,-3.5}$ (the color of the red giant branch at
$M_I$ = --3.5, corrected for reddening) and [Fe/H].  Using the distance
modulus derived above, $M_I$ = --3.5 corresponds to $I$ = 21.06 for And
VI.  Thus, the mean dereddened color for stars in the magnitude
interval 20.96 $\le$ $I$ $\le$ 21.16 and within 370 pixels (73 arcsec)
of And VI's center was computed, excluding one obvious non-member, and
yielding ($V$--$I$)$_{0,-3.5}$ = 1.38  $\pm$ 0.04 (where the error also
includes a 0.03 mag contribution due to photometric zeropoint
uncertainty and a 0.01 mag contribution from reddening uncertainty).
Applying the Armandroff et al.\ (1993) calibration results in a mean
abundance for And VI of $<$[Fe/H]$>$ = --1.58 $\pm$ 0.2.  This
metallicity is normal for a dwarf spheroidal (e.g., see Figure 9 of
Armandroff et al.\ 1993).  The mean metal abundance derived here for
And VI, --1.58 $\pm$ 0.2, agrees with that of Grebel \& Guhathakurta
(1999), --1.35 $\pm$ 0.3, to within the combined uncertainties.

The And VI giant branch in Figure \ref{cmd_fiducial} exhibits some
width in color at constant magnitude.  The standard deviation in
$V$--$I$ of the 24 And VI red giants used to calculate
($V$--$I$)$_{0,-3.5}$ is 0.09 mag.  This is much larger than the mean
photometric uncertainty for these same stars of 0.026 mag.  Although
this photometric uncertainty results from {\sc allstar} and does not
account for the crowding in the realistic way that artificial star
tests would, it does suggest that And VI exhibits a range of metal
abundance.  A standard deviation of 0.08 mag in $V$--$I$ at $M_I$ =
--3.5 corresponds to a standard deviation in abundance of 0.3 dex.
Grebel \& Guhathakurta (1999) also concluded that And VI has a range in
metal abundance.  A much more definitive investigation of the abundance
distribution in And VI will result from HST/WFPC2 observations to take
place during Cycle 8 (see the analogous results on the abundance spread
in And I by Da Costa et al.\ 1996).

The presence of upper asymptotic giant branch (AGB) stars, stars
significantly more luminous than the red giant branch tip, in
populations with [Fe/H] $\lesssim$ --1.0, is the signature of an
intermediate-age population (one whose age is considerably less than
that of the Galactic globular clusters).  Armandroff et al.\ (1993)
estimated the fraction of And I's \& And III's total luminosity that is
produced by an intermediate-age population using the numbers and
luminosities of upper-AGB stars, following the formalism developed by
Renzini \& Buzzoni (1986). This analysis yielded an intermediate-age
(3--10 Gyr) population fraction of 10 $\pm$ 10\% for both And I \&
III.  Following the same methodology, Paper I found no significant
evidence for upper-AGB stars, and thus no significant intermediate-age
population, in And V.

For And VI, we again adopt from Armandroff et al.\ (1993) the $M_{\rm
bol}$ limits that correspond to 3--10 Gyr upper AGB stars: --3.8 to
--4.6.  We use the bolometric corrections to $I$ magnitudes from
Da~Costa \& Armandroff (1990) and our distance modulus from the red
giant branch tip to calculate bolometric magnitudes.  In our And VI
($I$,$V$--$I$) diagram for stars within 73 arcsec of the And VI center
(Figure \ref{cmd_fiducial}), there are five stars within the above
$M_{\rm bol}$ interval (all in the range --3.8 to --4.0, with a mean
$M_{\rm bol}$ of --3.9; these are the stars above the red giant branch
tip with $I \approx 20.35$).  There are no such stars in the
color--magnitude diagram for objects with radius greater than 191
arcsec that we are using to estimate background (as discussed above).
Based on the surface brightness profile from Caldwell (1999), the area
from which we have selected upper-AGB stars represents 43\% of And VI's
total $M_V$ of --11.3.  Then Eq.\ 14 of Renzini \& Buzzoni (1986) and
the adoption of the AGB evolutionary rate used in Armandroff et
al.\ (1993) yields an intermediate-age (3--10 Gyr) population fraction
of 13 $\pm$ 6\% for And VI.  The error is estimated solely from
counting statistics for the upper-AGB stars.  Thus, And VI resembles
And I \& III in the strength of its intermediate-age population.  It
does not resemble the Galactic dwarf spheroidals Fornax or Leo I that
have dominant intermediate-age populations.

No evidence for young, blue main-sequence stars is seen in Figure
\ref{cmd_fiducial}.  Except for the star with $V$--$I$=0.13 and
$I$=21.30 that we identified as a candidate post-AGB star above, the
And VI ($I$,$V$--$I$) color--magnitude diagram is essentially devoid of
blue stars.  Bluer colors provide enhanced sensitivity to young
main-sequence stars.  Consequently, our ($V$,$B$--$V$) color--magnitude
diagram for And VI is displayed in Figure \ref{cmd_blue}, again for
stars within 73 arcsec of the And VI center.  Also plotted are red
giant branch fiducials for the Galactic globular clusters M15 ([Fe/H] =
--2.17) and NGC 1851 ([Fe/H] = --1.16) (Sarajedini \& Layden 1997)
shifted to the distance modulus and reddening of And VI.  As seen in
the ($I$,$V$--$I$) color--magnitude diagram, most of the stars in the
($V$,$B$--$V$) color--magnitude diagram are red giants that are
bracketed by the M15 and NGC 1851 fiducials.  Because our photometry is
significantly more precise in $V$ and $I$ than it is in $B$, we
estimate And VI's metallicity solely from the ($I$,$V$--$I$)
color--magnitude diagram.

One also sees a relatively small population of blue stars in Figure
\ref{cmd_blue}, with 23 $\lesssim$ $V$ $\lesssim$ 25.4 and -0.4
$\lesssim$ ($B$--$V$) $\lesssim$ 0.4.  These blue stars seen in the
($V$,$B$--$V$) diagram correspond to the medium brightness blue stars
apparent in our color image of And VI displayed in Figure
\ref{and6_ccd_image}.  Consider the 10 stars with 23.0 $\leq$ $V$
$\leq$ 24.0 and -0.1 $\leq$ ($B$--$V$) $\leq$ 0.4 in Figure
\ref{cmd_blue}.  The spatial distribution of such stars is consistent
with membership in And VI; their number density is an order of
magnitude lower in the corners of our images than it is within 73
arcsec of the And VI center.  In addition, the predictions of Galactic
star count models (e.g., Ratnatunga \& Bahcall 1985) for blue stars in
this magnitude range are two orders of magnitude smaller than
observed.  Therefore, we conclude that these blue stars are a component
of And VI.  Their distribution in the ($V$,$B$--$V$) diagram is not
indicative of main-sequence stars because they are not sufficiently
blue and because they fail to follow a continuous color--magnitude
relation.  Color--magnitude diagrams of the Galactic dSphs Leo I and
Sextans show populations of stars with very similar absolute magnitudes
and colors (Lee et al.\ 1993b; Demers et al.\ 1994; Gallart et
al.\ 1999; Mateo et al.\ 1995).  Some of these stars are known to be
anomalous Cepheids.  The HST/WFPC2 imaging of And VI planned for Cycle
8 should provide valuable clarification on this issue, because it will
reach substantially deeper with better precision than the current
photometry and should determine whether these blue stars are variable.
Based on the available data, And VI's color--magnitude diagrams are
consistent with those of other dwarf spheroidal galaxies, confirming
the other evidence for a dwarf spheroidal classification.
\placefigure{cmd_blue}

\section{Discussion}
The recent discoveries of And VI, And V (Paper I), and the Cas Dwarf
(Karachentsev \& Karachentseva 1999; Grebel \& Guhathakurta 1999),
coupled with the recognition that these dSph galaxies are part of the
extended M31 satellite system, have changed our view of this system.

Karachentsev (1996) discussed the spatial distribution of the
companions to M31.  The discoveries of And VI, And V and Cas modify the
spatial distribution of the M31 satellites.  One asymmetry in the
spatial distribution of the satellites results from And I, II \& III
all being located south of M31, while the three more luminous dwarf
elliptical companions NGC 147, 185 and 205 are all positioned north of
M31.  Also, Karachentsev (1996) noted that there are more M31
companions overall south of M31 than north of M31.  As discussed in
Paper I, the location of And V north of M31 lessens both of these
asymmetries, and Cas's location north of M31 reduces the asymmetries
further (see Figure 2). Because And VI is south of M31, it diminishes
the asymmetry reduction provided by the Cas Dwarf.

\begin{figure}[p]
\epsscale{1.0}
\plotone{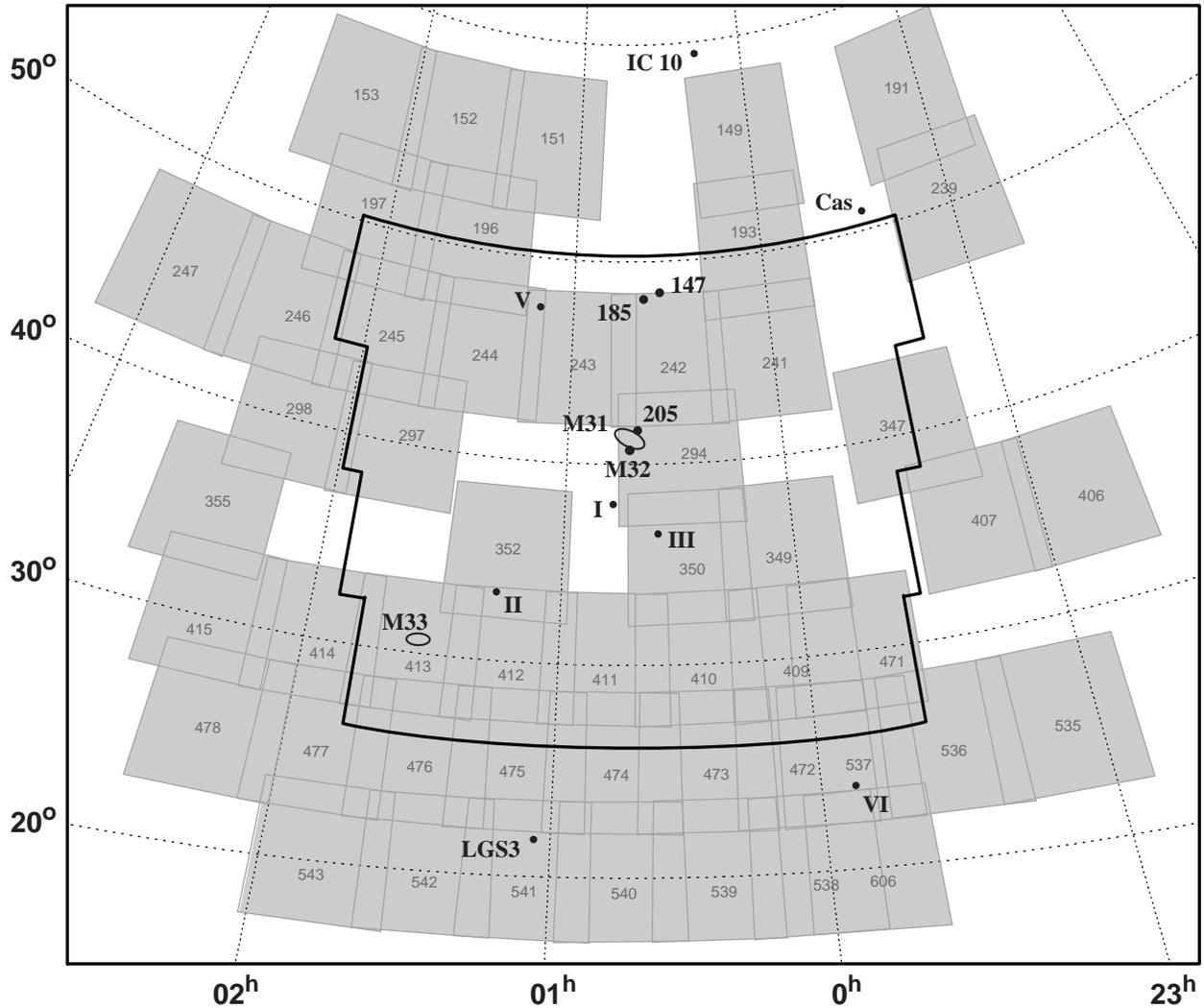}
\caption[]{Map of the region of sky around M31 that has been surveyed
for low surface brightness dwarf galaxies.  Individual POSS-II plates
that have been searched are shown in grey and labeled; these cover
$\sim$1550 deg$^2$.  Our effort to survey the remaining area in this
diagram is ongoing.  The area surveyed by van den Bergh (1972, 1974) is
outlined in bold.  M31 and its well known neighbors are labeled.  The
new M31 dwarf spheroidals, And VI, And V (Armandroff et al.\ 1998b) and
Cas (Karachentsev \& Karachentseva 1999; Grebel \& Guhathakurta 1999),
are also indicated.}
\label{survey_map}
\end{figure}

Karachentsev (1996) called attention to the elongated shape of the
distribution of M31 companions, with axis ratio 5:2:1 (based on the
galaxies labelled in Figure \ref{survey_map} that were known in 1996
plus the more distant galaxy IC 1613, all of which he argues are
associated with M31).  On the sky, this elongation occurs primarily
along the north--south axis.  As discussed in Paper I, And V is located
within this flattened spatial distribution.  Of all the M31 companions,
And VI's location to the southeast of M31 places it the furthest from
this elongated distribution, and nearly on the distribution's minor
axis.  The Cas Dwarf is the next most deviant companion from the
elongated pattern.

With projected radii from the center of M31 of 112, 271 and 224 kpc,
respectively, And V, And VI and Cas increase the mean projected radius
of the M31 dwarf spheroidals from 86 kpc to 144 kpc.  Karachentsev
(1996) also called attention to morphological segregation among the M31
companions, with the dwarf ellipticals and dwarf spheroidals located
closer to M31, and the spirals and irregulars on the periphery (see
also van den Bergh 1994 regarding segregation among the Galactic
companions).  The identification of And VI and Cas extends the presence
of dwarf spheroidals to galactocentric distances comparable to those of
LGS 3 (a transition object between dwarf irregular and dwarf
spheroidal) and IC 10 (a dwarf irregular).  Thus, And VI and Cas weaken
the case for morphological segregation in the M31 satellite system.

The discovery of dwarf galaxies like And VI, And V and Cas augments the
faint end of the luminosity function of the Local Group.  Because of
the special way that nearby dwarf galaxies are selected, the Local
Group is uniquely suited to the study of the faint end of the galaxy
luminosity function (e.g., Pritchet \& van den Bergh 1999).  Accurate
integrated $V$ magnitudes are now available for And VI, And V and Cas
from wide-field CCD imaging (Caldwell 1999).  When combined with the
distance moduli (this paper; Paper I; Grebel \& Guhathakurta 1999),
these yield $M_V$ values of -11.3, -9.1, and -12.0 for And VI, And V,
and Cas, respectively.  Luminosity functions for M31's dwarf companions
and for the entire Local Group that include and highlight these three
new dwarf spheroidals are shown in Figure \ref{new_gal_lum_fun}.  The
Local Group and subgroup memberships and $M_V$ values for the galaxies
besides And VI, And V, and Cas are from Mateo (1998), except for the
LMC and SMC which are from Westerlund (1997).  And VI, And V and Cas
increase the number of galaxies with -9.0 $>$ $M_V$ $\geq$ -12.0 that
are M31 companions by 75\% and in the entire Local Group by 19\%.
\placefigure{new_gal_lum_fun}

From a survey of nine clusters of galaxies, Trentham (1998) derived a
composite luminosity function that is steeper at the faint end than
that of the Local Group (see his Figure 2)\@.  He attributed the
seeming shortfall at faint $M_V$ in the Local Group to poor counting
statistics and/or incompleteness.  The discovery of And VI, And V and
Cas reduces somewhat the difference between the Local Group luminosity
function and Trentham's (1998) function.  To illustrate the difference,
we have plotted a version of the Trentham luminosity function,
transformed from $B$ to $V$ assuming ($B$--$V$)$_0$ = 0.70, in Figure
\ref{new_gal_lum_fun}.  The Trentham function has been normalized to
match the M31 and Local Group functions in the interval -16 $>$ $M_V$
$\geq$ -18.  One sees that the luminosity function for clusters of
galaxies contains many more galaxies for $M_V$ $\gtrsim$ -13 than are
known in the Local Group.  Despite the inclusion of the three newly
discovered Local Group dwarf spheroidals, the difference between the
Trentham (1998) luminosity function and that of the Local Group remains
significant at faint $M_V$ ($>$ 95\% confidence level).  Pritchet \&
van den Bergh (1999) reached the same conclusion using a somewhat
different assemblage of Local Group data.  Provided that the Trentham
function is representative, we conclude that either faint dwarf
galaxies are much less common in the Local Group than in richer
clusters or that the Local Group sample is incomplete by $\sim$10 to
$\sim$25 galaxies in the $M_V$ range --12 to --14.

One important issue remains regarding Figure \ref{new_gal_lum_fun}.  Is
the peak of the histogram in the region -11 $\gtrsim$ $M_V$ $\gtrsim$
-12, followed by a decline at fainter magnitudes, real or is it caused
by incompleteness and small number statistics?  The luminosity function
is expected to decline fainter than some $M_V$ because of the inability
of sufficiently small proto-galactic gas clouds to collapse and form
stars due to photoionization by ultraviolet background radiation (e.g.,
Thoul \& Weinberg 1996).  Our continuing search for faint dwarf
galaxies in the vicinity of M31 and other Local Group searches should
help resolve the issue of where this turnover occurs.

Beyond the observational evidence for a deficit of dwarfs in the Local
Group based on Trentham's (1998) luminosity function, similarly
dramatic discrepancies (factors of 15 or more) are suggested by recent
theoretical developments.  Klypin et al.\ (1999), for example, present
simulations of the hierarchical formation of small groups like our
Local Group in the currently favored $\Lambda$CDM universe.  While
$\sim$40 Local Group galaxies are known, those authors predict that
$\sim$500 galaxies should exist for which the vast majority are low
luminosity dwarfs.  That number doubles when the search radius extends
from $\sim$300 to $\sim$600 kpc from the host galaxy.  Our continuing
survey around M31, especially when we extend it to larger radii, will
provide a tight constraint on the true number of faint dwarf galaxies
in the Local Group.

\acknowledgments
The Digitized Sky Surveys were produced at the Space Telescope Science
Institute under U.S.\ Government grant NAG W-2166.  The survey images
used here are based on photographic data obtained using the Oschin
Schmidt Telescope on Palomar Mountain.  The Second Palomar Observatory
Sky Survey was made by the California Institute of Technology with
funds from the National Science Foundation, the National Geographic
Society, the Sloan Foundation, the Samuel Oschin Foundation, and the
Eastman Kodak Corporation.  The Oschin Schmidt Telescope is operated by
the California Institute of Technology and Palomar Observatory.
Supplemental funding for sky-survey work at the STScI is provided by
the European Southern Observatory.  Watanabe Masaru at the National
Astronomical Observatory of Japan and Jesse Doggett and especially the
late Barry Lasker at STScI kindly helped us to access the DSS data.  We
appreciate the efforts of the WIYN Queue Team, Dianne Harmer, Abi Saha,
Paul Smith, and Daryl Willmarth, in obtaining the WIYN images.  We are
grateful to Nelson Caldwell, Gary Da Costa, Eva Grebel, Anatoly Klypin,
Chris Pritchet, and Abi Saha for helpful discussions.  J.\ E.\ D.\ was
supported by the NSF Research Experiences for Undergraduates Program at
NOAO during the summer of 1997.  G.\ H.\ J.\ wishes to thank Peter
Strittmatter for providing a sabbatical office at Steward Observatory
where some of this work was done.  J.\ E.\ D.\ wishes to thank Robert
Williams for providing an office at STScI where some of this work was
done.

\begin{deluxetable}{lccccc}
\tablecolumns{5}
\tablecaption{Log of And VI WIYN Imaging}
\tablewidth{0pt}
\tablehead{
\colhead{Description}& \colhead{Date
(U.T.)}&\colhead{Filter}&\colhead{Exp.\ Time}&
\colhead{FWHM ($^{\prime\prime}$)}}
\startdata
And VI deep&1998 Oct 19&$B$&2 $\times$ 2000 s&0.8&\nl
And VI deep&1998 Oct 14&$V$&3 $\times$ 1000 s&0.6&\nl
And VI deep&1998 Oct 15&$I$&2 $\times$ 1000 s&0.7&\nl
And VI calibration&1998 Sept 20&$B$&600 s&1.2\nl
And VI calibration&1998 Sept 20&$V$&1000 s&1.1\nl
And VI calibration&1998 Sept 20&$I$&300 s&1.0\nl
\enddata
\label{obs_table}
\end{deluxetable}
\newpage

\begin{figure}[p]
\caption[]{The left panel shows an  image of And VI from the digitized
POSS-II (IIIaF + RG610 plate/filter combination).  The right panel
shows the effect of applying our digital enhancement procedure to the
POSS-II image. Each panel is 17 arcmin on a side.  North is at the top,
and east is to the left.  Note the similarities between And VI in these
images and And II \& III in the analogous Figure 1 of Armandroff et
al.\ (1998b).}
\label{and6dss}
\end{figure}

\begin{figure}[p]
\caption[]{Color image of And VI made from $B$, $V$, and $I$ exposures
from the WIYN 3.5-m telescope.  The size of the image is $148 \times
217$ arcsec.  North is at the top, and east is to the left.}
\label{and6_ccd_image}
\end{figure}

\begin{figure}[p]
\caption[]{An H$\alpha$ image of And VI made from the combination of
three 700-second exposures with the KPNO 4-m telescope is shown in the
left panel.  A continuum-subtracted version of this image is displayed
in the right panel.  In each panel, north is at the top, and east is to
the left.  Each image is 198 arcsec $\times$ 242 arcsec in size.  The
residuals in the subtracted image surrounding the brighter stars are
due to small differences in the PSF between $R$ and H$\alpha$.  Note
the lack of diffuse H$\alpha$ emission or H {\sc ii} regions.}
\label{and6halpha}
\end{figure}

\begin{figure}[p]
\epsscale{0.90}
\plotone{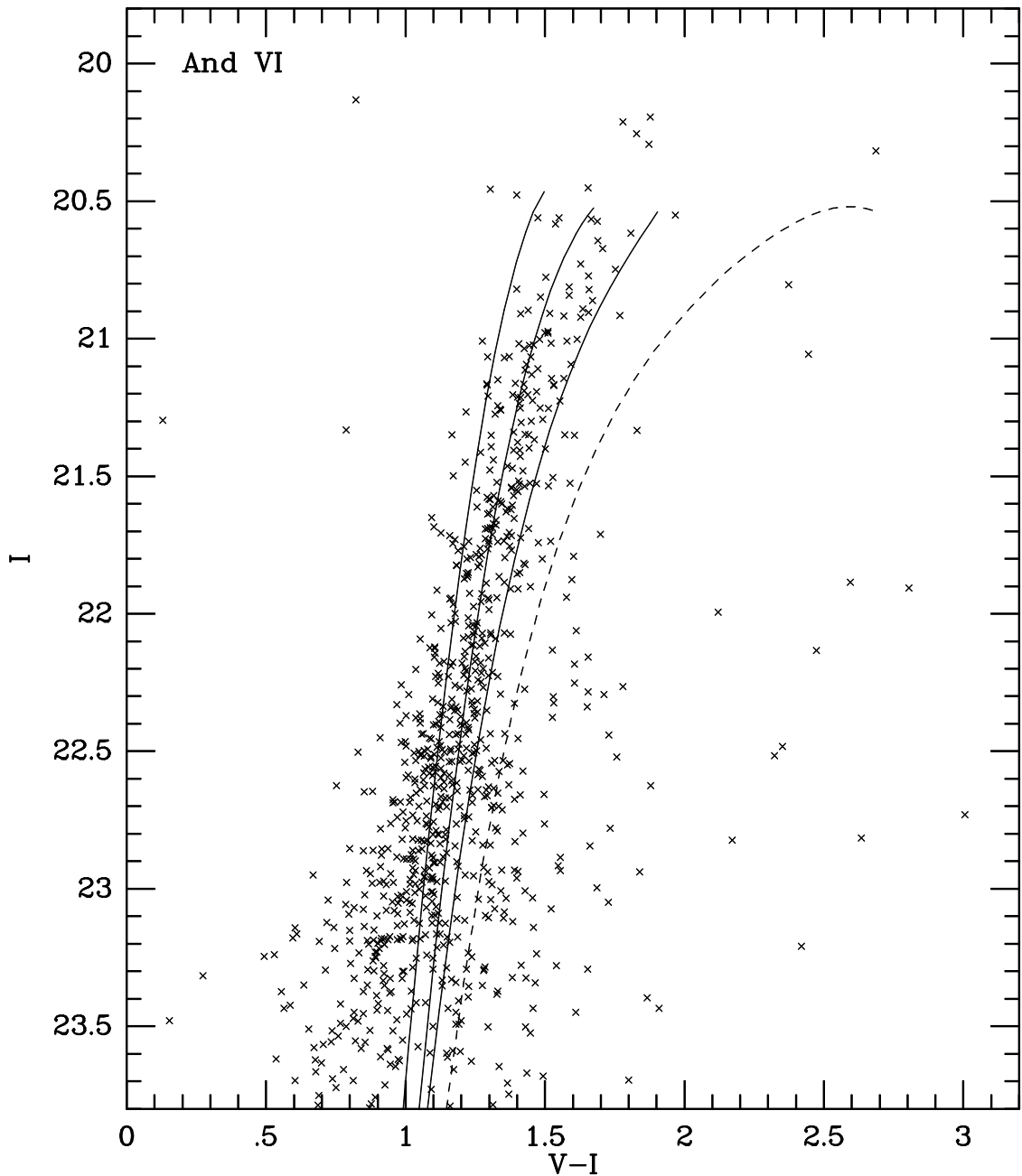}
\caption[]{Color--magnitude diagram in $I$ vs $V$--$I$ for stars within
a radius of 370 pixels (73 arcsec) of the center of And VI, where And
VI members greatly outnumber field contamination.  Red giant branch
fiducials for four Galactic globular clusters that span a range of
metal abundance (Da Costa \& Armandroff 1990), shifted to the distance
modulus and reddening of And VI, have been overplotted.  From left to
right, the red giant branch fiducials are M15 ([Fe/H] = --2.17), M2
(--1.58), NGC 1851 (--1.16), and 47 Tuc (--0.71, dashed line).}
\label{cmd_fiducial}
\end{figure}

\begin{figure}[p]
\plotone{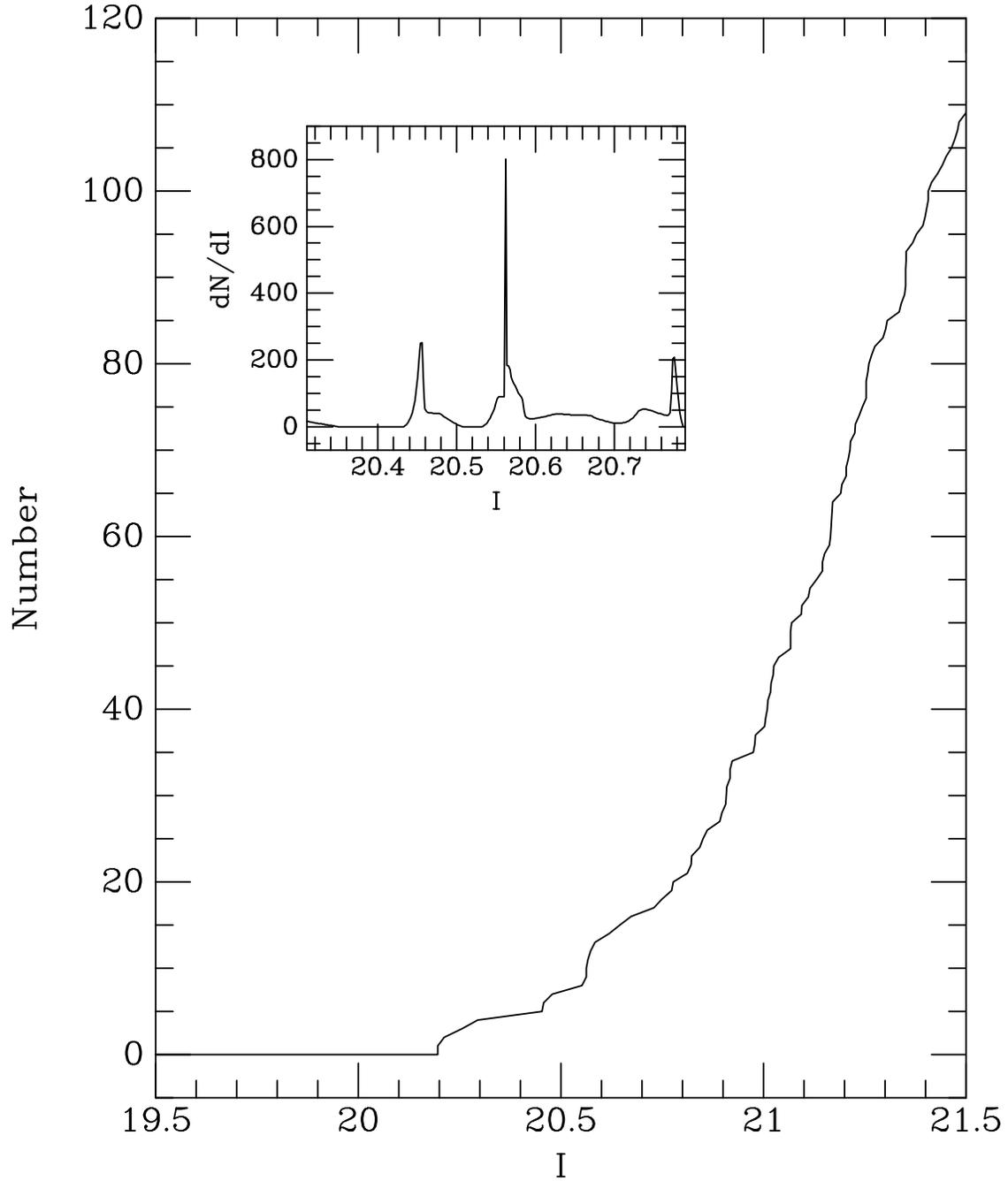}
\caption[]{Cumulative luminosity function in $I$ for stars within a
radius of 370 pixels (73 arcsec) of the center of And VI, where And VI
members greatly outnumber field contamination.  The inset shows the
derivative of the cumulative luminosity function in the magnitude range
near the red giant branch tip.  The strong rise in the derivative at
$I$ = 20.56 indicates the location of the red giant branch tip.  The
modest rise at $I$ = 20.46 may be due to the presence of AGB stars.}
\label{and6_lf}
\end{figure}

\begin{figure}[p]
\plotone{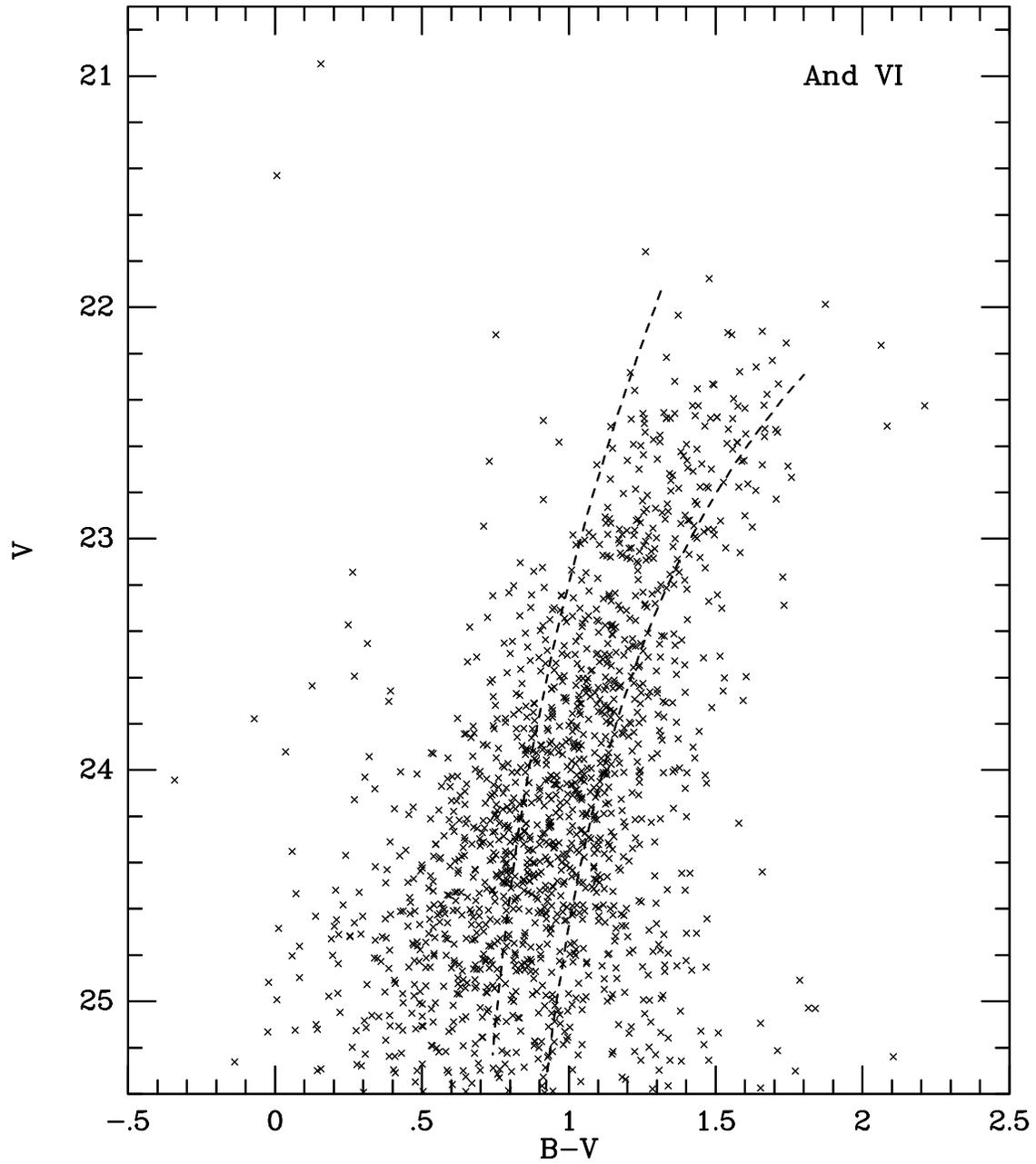}
\caption[]{Color--magnitude diagram in $V$ vs $B$--$V$ for stars within
a radius of 370 pixels (73 arcsec) of the center of And VI, where And
VI members greatly outnumber field contamination.  The dashed lines are
red giant branch fiducials for the Galactic globular clusters M15
([Fe/H] = --2.17; left curve) and NGC 1851 ([Fe/H] = --1.16; right
curve) from Sarajedini \& Layden (1997), shifted to the And VI distance
modulus and reddening.}
\label{cmd_blue}
\end{figure}

\begin{figure}[p]
\epsscale{0.85}
\plotone{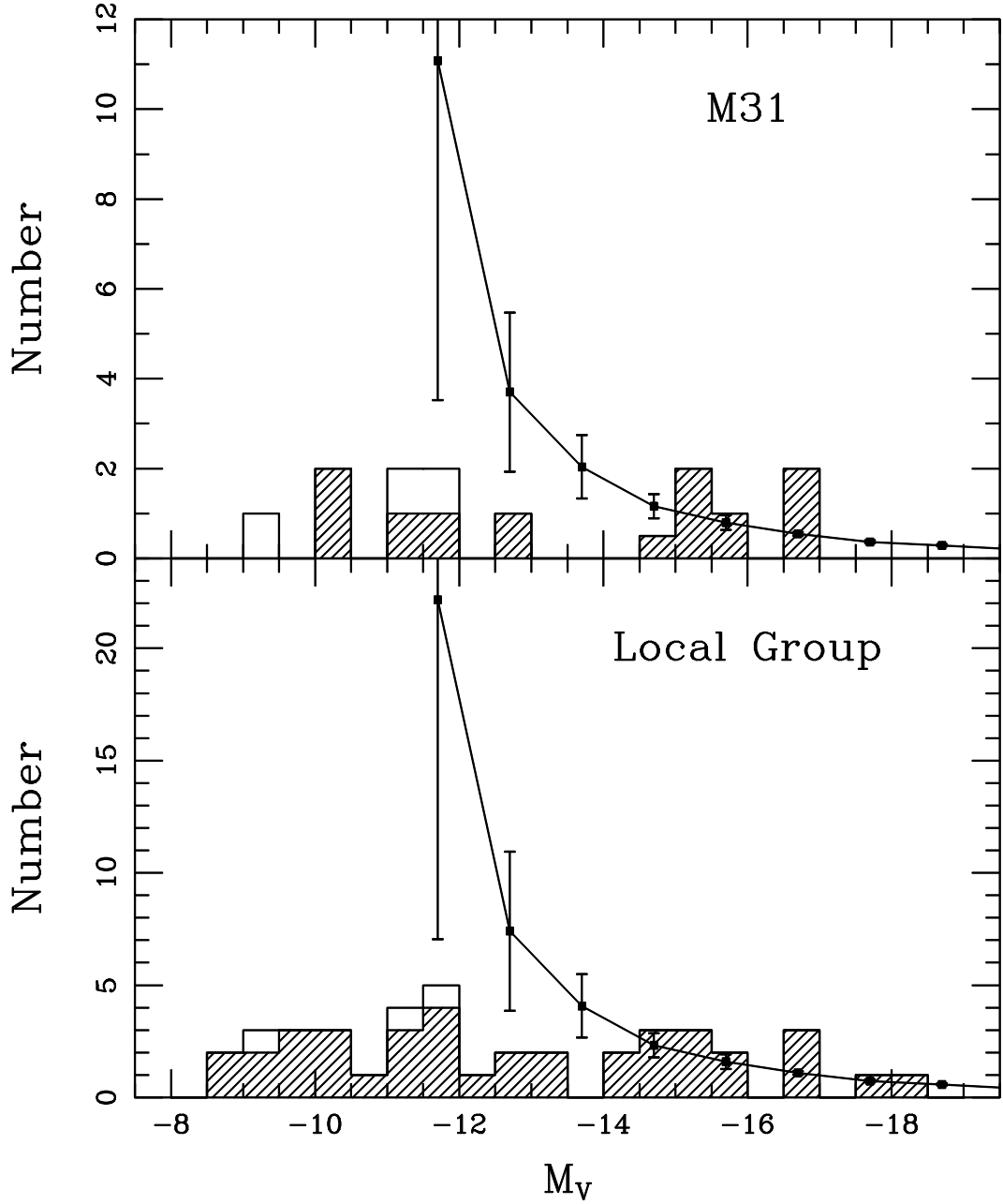}
\caption[]{Luminosity functions for dwarf galaxies associated with M31
(upper panel) and the entire Local Group (lower panel).  Galaxies known
before our survey began are indicated by shaded histograms, while the
three new M31 dwarf spheroidals are shown as open histograms.  The
continuous function in each panel is the composite luminosity function
for nine clusters of galaxies from Trentham (1998), transformed from
$B$ to $V$.  Note that the positive error bar on the faintest data
point from Trentham is off scale but symmetric with the negative error
bar.}
\label{new_gal_lum_fun}
\end{figure}

\end{document}